\documentstyle[11pt,fleqn]{article}

\begin{document}

\title{{\bf Analytic representations based on 
SU(1,1) coherent states and their applications}}
\author{ {\Large
C Brif,${}^{1}$\thanks{E-mail: costya@physics.technion.ac.il;
URL address: http://www.technion.ac.il/\~{}brif} 
\ A Vourdas${}^{2}$\thanks{E-mail: ee21@liverpool.ac.uk} 
\ and A Mann${}^{1}$\thanks{E-mail: ady@physics.technion.ac.il} } \\
\\ 
{\normalsize ${}^{1}$Department of Physics, Technion -- 
Israel Institute of Technology, Haifa 32000, Israel } \\ 
{\normalsize ${}^{2}$Department of Electrical Engineering and 
Electronics, University of Liverpool, }  \\
{\normalsize Brownlow Hill, Liverpool L69 3BX, United Kingdom} }
\date{}
\maketitle

%{\large
%\vspace*{0.3cm}
%Short title: SU(1,1) analytic representations

%\vspace*{0.3cm}
%PACS numbers: 02.20.Qs, 03.65.Fd, 42.50.Dv
%}

%\newpage
%\setlength{\baselineskip}{0.73cm}

\begin{abstract}
%{\large
\noindent
We consider two analytic representations of the SU(1,1) Lie group:
the representation in the unit disk based on the SU(1,1) Perelomov
coherent states and the Barut-Girardello representation based on
the eigenstates of the SU(1,1) lowering generator. We show that
these representations are related through a Laplace transform. 
A ``weak'' resolution of the identity in terms of the Perelomov
SU(1,1) coherent states is presented which is valid
even when the Bargmann index $k$ is smaller than one half.
Various applications of these results in
the context of the two-photon realization of SU(1,1)
in quantum optics are also discussed. 
%}
\end{abstract}

%\newpage

\section{Introduction}
\setcounter{equation}{0}

\noindent
Analytic representations based on overcomplete sets of quantum states 
are a mathematical tool used frequently in quantum optics and other 
areas of quantum physics. These representations have also an important
physical meaning as they provide a natural description of the
quantum-classical correspondence. The most familiar example of such
a representation is the Bargmann analytic representation (Bargmann 1961)
based on the overcomplete set of the Glauber coherent states 
(Glauber 1963). This representation has also been studied by Fock (1928) 
and Segal (1962).

We focus here on two well-known analytic representations of 
the SU(1,1) Lie group: one is the analytic representation 
in the unit disk based on the overcomplete set of the SU(1,1) 
Perelomov coherent states (Perelomov 1972, 1977, 1986); 
and the other is the Barut-Girardello representation based on 
the overcomplete basis of the Barut-Girardello states 
(Barut and Girardello 1971).
These representations are useful in many quantum mechanical problems 
involving dynamical systems with SU(1,1) symmetry (Kuriyan {\em et al} 
1967, Sharma {\em et al} 1978, Mukunda {\em et al} 1980, Sharma {\em 
et al} 1981, Kim and Noz 1986).
The analytic representation in the unit disk is also related to
Berezin's quantization on homogeneous K\"{a}hlerian manifolds 
(Berezin 1974, 1975a, b, Perelomov 1986, Bar-Moshe and Marinov 1994). 
Some other properties and applications of the analytic 
representation in the unit disk were studied by Gerry (1983), 
Sudarshan (1993), Vourdas (1992, 1993a, b) and by Brif and Ben-Aryeh 
(1994b, 1996). 
The Barut-Girardello representation was used by Basu (1992),
Trifonov (1994) and by Prakash and Agarwal (1995).
 
In the present work we study some interesting properties, 
relations and applications of these two representations. 
We show that the Barut-Girardello representation and the analytic 
representation in the unit disk are related through a Laplace 
transform. We also consider a resolution of the identity 
in terms of the SU(1,1) coherent states
which is known to be valid only for $k>1/2$. Using an analytic 
continuation we introduce a ``weak'' resolution of the identity
which is valid even for $k<1/2$.
We apply this method to the two-photon realization 
of the SU(1,1) Lie algebra in quantum optics, and we obtain the 
squeezed-state analytic representation. In this context we show that 
a synthesis of the Barut-Girardello $k=1/4$ and $k=3/4$ representations 
is related to the Bargmann analytic representation.
We also construct analytic representations based on displaced
squeezed states and use them to study the energy eigenstates of the 
squeezed and displaced harmonic oscillator.

\section{SU(1,1) analytic representations}

\noindent
The group SU(1,1) is the most elementary noncompact non-Abelian simple 
Lie group. It has several series of unitary irreducible representations: 
discrete, continuous and supplementary (Bargmann 1947; Vilenkin 
1968). The Lie algebra corresponding to the group SU(1,1) is spanned
by the three group generators $\{K_{1},K_{2},K_{3}\}$,
	\begin{equation}
[K_{1} , K_{2}] = - {\rm i} K_{3}  \;\;\;\;\;\;\;\; 
[K_{2} , K_{3}] = {\rm i} K_{1}  \;\;\;\;\;\;\;\;
[K_{3} , K_{1}] = {\rm i} K_{2} .  \label{2.0.1}
	\end{equation}
It is customary to use the raising and lowering generators 
$K_{\pm} = K_{1} \pm {\rm i} K_{2}$, which satisfy
	\begin{equation}
[K_{3} , K_{\pm}] = \pm K_{\pm}  \;\;\;\;\;\;\;\;\; 
[K_{-} , K_{+}] = 2K_{3}. \label{2.0.2}
	\end{equation}
The Casimir operator
$K^{2} = K_{3}^{2} -K_{1}^{2} -K_{2}^{2}\,$ 
for any irreducible representation is
$K^{2} = k(k-1) I$. 
Thus a representation of SU(1,1) is determined by the number $k$. 
The corresponding Hilbert space ${\cal H}_{k}$ is spanned 
by the complete orthonormal basis $|n,k\rangle$ $(n=0,1,2,\ldots)$:
	\begin{equation}
\langle m,k|n,k \rangle = \delta_{mn}  \;\;\;\;\;\;\;\;\; 
\sum_{ n = 0 }^{\infty} |n,k \rangle\langle n,k| = I .  
\label{2.0.3}
	\end{equation}
Various sets of states can be defined in the representation Hilbert
space. The overcomplete bases of the SU(1,1) coherent states and of 
the Barut-Girardello states
are of special importance because of their remarkable mathematical
properties and interesting physical applications.

\subsection{Analytic representations in the unit disk}

\noindent
As was discussed by Perelomov (1972, 1977, 1986),
each SU(1,1) coherent state corresponds to a point in the coset
space SU(1,1)/U(1) that is the upper sheet of the two-sheet 
hyperboloid (Lobachevski plane). Thus a coherent state is specified 
by a pseudo-Euclidean unit vector of the form 
$(\sinh\tau \cos\varphi , \sinh\tau \sin\varphi , \cosh\tau)$.
The coherent states $| \zeta,k \rangle$ are obtained by applying 
unitary operators 
$\Omega(\xi) \in$ SU(1,1)/U(1) to the lowest state $|n=0,k \rangle$,
	\begin{eqnarray}
| \zeta,k \rangle & = & \exp \left( \xi K_{+} -\, \xi^{\ast} 
K_{-} \right) | 0,k \rangle  = (1-|\zeta|^{2})^{k} \exp 
\left( \zeta K_{+} \right) |0,k \rangle   \nonumber \\
& = & (1-|\zeta|^{2})^{k} \sum_{n = 0}^{\infty} 
\left[\frac{\Gamma(n+2k)}{n!\Gamma(2k)}\right]^{1/2}\!\zeta^{n} 
|n,k \rangle .         \label{2.1.1}
	\end{eqnarray}
Here $\xi=-(\tau/2)\, {\rm e}^{- {\rm i} \varphi}$ and 
$\zeta = (\xi/|\xi|) \tanh |\xi| = -\tanh (\tau/2)\, 
{\rm e}^{- {\rm i} \varphi}$, so $|\zeta| < 1$. 
The condition $|\zeta|<1$ shows that the 
SU(1,1) coherent states are defined in the interior 
of the unit disk.
An important property is the resolution of the identity:
for $k>1/2$ one gets
	\begin{equation}
\int {\rm d} \mu (\zeta,k)  | \zeta,k \rangle \langle \zeta,k | = I   
\mbox{\hspace{1.5cm}}
{\rm d} \mu (\zeta,k) = \frac{2k-1}{\pi} \frac{ {\rm d}^{2} 
\zeta}{(1-|\zeta|^{2})^{2}}        \label{2.1.2}
	\end{equation}
where the integration is over the unit disk $|\zeta|<1$. 
For $k=\frac{1}{2}$ the limit $k\rightarrow \frac{1}{2}$ must be 
taken after the integration is carried out in the general form. 

One can represent the state space ${\cal H}_{k}$ as the Hilbert 
space of analytic functions $G(\zeta;k)$ in the unit disk 
${\cal D}(|\zeta|<1)$. They form the so-called Hardy space 
$H_{2}({\cal D})$. For a normalized state
$| \Psi \rangle = \sum_{n = 0}^{\infty} C_{n} |n,k \rangle$, 
one gets 
	\begin{equation}
G(\zeta;k) = (1-|\zeta|^{2})^{-k} \langle \zeta^{\ast},k |\Psi\rangle 
= \sum_{n = 0}^{\infty} C_{n} \left[\frac{\Gamma(n+2k)
}{n!\Gamma(2k)}\right]^{1/2}\!\zeta^{n}   
\label{2.1.3}
	\end{equation}
	\begin{equation}
| \Psi \rangle = \int {\rm d} \mu (\zeta,k)  (1-|\zeta|^{2})^{k} 
G(\zeta^{\ast};k) | \zeta,k \rangle   \label{2.1.4}
	\end{equation}
and the scalar product is 
 	\begin{equation}
\langle \Psi_{1} | \Psi_{2} \rangle = \int {\rm d} \mu (\zeta,k)  
(1-|\zeta|^{2})^{2k} [G_{1}(\zeta;k)]^{\ast} G_{2}(\zeta;k) 
	\end{equation}
This is the analytic representation in the unit disk. The generators 
$K_{\pm}$ and $K_{3}$ act on the Hilbert space of entire functions 
$G(\zeta;k)$ as first-order differential operators:
	\begin{equation}
K_{+} =  \zeta^{2} \frac{ {\rm d} }{ {\rm d} \zeta} + 2k\zeta  
\;\;\;\;\;\;\;\;
K_{-} = \frac{ {\rm d} }{ {\rm d} \zeta}    \;\;\;\;\;\;\;\;
K_{3} =  \zeta \frac{ {\rm d} }{ {\rm d} \zeta} + k  .   \label{2.1.5}
	\end{equation}

The SU(1,1) transformations are easily implemented in this 
representation through M\"{o}bius conformal mappings:
	\begin{equation}
\zeta \rightarrow \frac{a\zeta + b}{b^{\ast}\zeta + a^{\ast}}
\;\;\;\;\;\;\;\;\;\; |a|^2 - |b|^2 = 1 .  \label{2.1.6}
	\end{equation}
The analytic function is then transformed as
	\begin{equation}
G(\zeta;k) \rightarrow G\left( \frac{a\zeta + b}{
b^{\ast}\zeta + a^{\ast}};
k \right) (b^{\ast}\zeta + a^{\ast})^{-2k} .    \label{2.1.7}
	\end{equation}
The relation between the parameters $a$, $b$ appearing here and the 
parameter $\xi=-(\tau/2)\, {\rm e}^{- {\rm i} \varphi}$ of equation 
(\ref{2.1.1}) is
	\begin{equation}
a = \cosh (\tau/2) \;\;\;\;\;\;\;\;\;\; b = \sinh (\tau/2) \, 
{\rm e}^{ {\rm i} \varphi} .              \label{2.1.8}
	\end{equation}

\subsection{The Barut-Girardello analytic representation}

\noindent
Barut and Girardello (1971) constructed the eigenstates of the 
lowering generator $K_{-}$,
	\begin{equation}
K_{-} |z,k\rangle = z |z,k\rangle        \label{2.2.1}
	\end{equation}
where $z$ is an arbitrary complex number. The Barut-Girardello states 
can be decomposed over the orthonormal state basis,
	\begin{equation}
|z.k\rangle = \frac{z^{k-1/2}}{\sqrt{I_{2k-1}(2|z|)}} 
\sum_{n = 0}^{\infty} \frac{z^{n}}{\sqrt{n!\Gamma(n+2k)}} 
|n,k\rangle     \label{2.2.2}
	\end{equation}
where $I_{\nu}(x)$ is the $\nu$-order modified Bessel function of 
the first kind. The Barut-Girardello states are normalized but they are 
not orthogonal to each other. Various properties of these states were
studied in different physical contexts, i.e. for different 
realizations of SU(1,1) (Dodonov {\em et al} 1974, Sharma {\em et al}
1978, Mukunda {\em et al} 1980, Sharma {\em et al} 1981, Agarwal 1988, 
Bu\v{z}ek 1990, Basu 1992, Brif and Ben-Aryeh 1994a, Brif 1995).
It is not difficult to prove the following identity resolution
(Barut and Girardello 1971):
	\begin{equation}
\int {\rm d} \mu(z,k) |z,k \rangle\langle z,k| = I 
\mbox{\hspace{1.5cm}}
{\rm d} \mu(z,k) = \frac{2}{\pi} K_{2k-1}(2|z|) I_{2k-1}(2|z|) 
{\rm d}^{2} z .
\label{2.2.3}
	\end{equation}
The integration in (\ref{2.2.3}) is over the whole $z$ plane, 
and $K_{\nu}(x)$ is the $\nu$-order modified Bessel function of the 
second kind. The identity resolution (\ref{2.2.3}) holds for $k>0$.
Thus the Barut-Girardello states form, for each positive $k$, an 
overcomplete basis in ${\cal H}_{k}$.

The state space ${\cal H}_{k}$ can be represented as the
Hilbert space of entire functions $F(z;k)$ which are analytic over 
the whole $z$ plane. For a normalized state $|\Psi\rangle$, we get
	\begin{equation}
F(z;k) = \frac{\sqrt{I_{2k-1}(2|z|)}}{z^{k-1/2}}
\langle z^{\ast},k|\Psi\rangle = \sum_{n = 0}^{\infty} 
\frac{C_{n} }{ \sqrt{n!\Gamma(n+2k)} } z^{n} 
\label{2.2.4}
	\end{equation}
	\begin{equation}
|\Psi\rangle = \int {\rm d} \mu(z,k) \frac{(z^{\ast})^{k-1/2}
}{\sqrt{I_{2k-1}(2|z|)}} F(z^{\ast};k) |z,k\rangle 
\label{2.2.5}
	\end{equation}
and the scalar product is
	\begin{equation}
\langle \Psi_1 | \Psi_2 \rangle = \int {\rm d} \mu(z,k) 
\frac{|z|^{2k-1}}{I_{2k-1}(2|z|)} [F_{1}(z;k)]^{\ast} F_{2}(z;k) .
	\end{equation}
The SU(1,1) generators $K_{\pm}$ and $K_{3}$ act on the Hilbert
space of entire functions $F(z;k)$ as linear operators:
	\begin{equation}
K_{+} =  z   \;\;\;\;\;\;\;\;
K_{-} = 2k \frac{ {\rm d} }{ {\rm d} z} + z \frac{ {\rm d}^{2} }{ 
{\rm d} z^{2}}  \;\;\;\;\;\;\;\;
K_{3}  = z \frac{ {\rm d} }{ {\rm d} z} + k  .   
\label{2.2.6}
	\end{equation}

\section{A Laplace transform between the two analytic representations}

\noindent
It can be easily seen that the unit-disk and Barut-Girardello 
analytic representations are closely related to each other. 
We find, for example, that the scalar product between the Perelomov
and Barut-Girardello states is
	\begin{equation}
\langle \zeta^{\ast},k|z,k \rangle = \left[ \frac{ z^{k-1/2}
(1-|\zeta|^2)^{k} }{ \sqrt{I_{2k-1}(2|z|) \Gamma(2k)} } \right]
\exp(z\zeta) .       \label{3.1}     
	\end{equation}
For convenience, we introduce the complex number $\rho \equiv 1/\zeta$ 
($|\rho|>1$). Then, according to equation (\ref{2.1.3}), we write
	\begin{equation}
G(1/\rho;k) = \sum_{n=0}^{\infty} C_{n} \left[ \frac{ \Gamma(n+2k) 
}{ n! \Gamma(2k) } \right]^{1/2} \frac{1}{\rho^{n}} .
\label{3.2} 
	\end{equation}
Then, by using the expression for the Gamma function,
	\begin{equation}
\Gamma(n+2k) = \rho^{n+2k} \int_{0}^{\infty} t^{n+2k-1}\, 
{\rm e}^{-\rho t} {\rm d} t \;\;\;\;\;\;\;\;\;\; {\rm Re}\, \rho > 0
	\end{equation}
we obtain the following result:
	\begin{equation}
G(1/\rho;k) = \frac{\rho^{2k}}{\sqrt{\Gamma(2k)}} \left\{
\int_{0}^{\infty} \left[ z^{2k-1} F(z;k) \right] {\rm e}^{-\rho z} 
{\rm d} z \right\} \;\;\;\;\;\;\;\;\;\; {\rm Re}\, \rho > 0 \;\;\;\;\;
|\rho| > 1 .    \label{3.3} 
	\end{equation}
This relation gives $G(\zeta;k)$ in the right half of the unit disk
(${\rm Re}\, \zeta > 0$), but since $G(\zeta;k)$ is analytic, equation 
(\ref{3.3}) effectively defines $G(\zeta;k)$ in the whole unit disk.
Therefore, we see that the two analytic representations are related
via the Laplace transform (Erd\'{e}lyi {\em et al} 1953b, ch IV). 
The inverse Laplace transform (Erd\'{e}lyi {\em et al} 1953b, ch V) 
yields, correspondingly,
	\begin{equation}
F(z;k) = \frac{\sqrt{\Gamma(2k)}}{z^{2k-1}} \left\{ 
\frac{1}{2\pi {\rm i} } \int_{1- {\rm i} \infty}^{1+ {\rm i} \infty} 
\left[ \rho^{-2k} G(1/\rho;k) \right]
{\rm e}^{\rho z} {\rm d} \rho \right\} .           \label{3.4}
	\end{equation}
Note that equations (\ref{3.3}) and (\ref{3.4}) can be also 
formally considered as the Mellin transform (Erd\'{e}lyi {\em et al} 
1953b, ch VI) and the inverse Mellin transform (Erd\'{e}lyi 
{\em et al} 1953b, ch VII), respectively.
Using the relation 
(\ref{3.4}), we readily find how the SU(1,1) transformations affect the 
Barut-Girardello representation. Let a normalized state $|\Psi\rangle$
be represented by the functions $G(\zeta;k)$ and $F(z;k)$. 
If the SU(1,1) transformations are implemented through the M\"{o}bius
conformal mappings (\ref{2.1.6}), the function $G(\zeta;k)$ is 
transformed according to (\ref{2.1.7}). Then the relation (\ref{3.4}) 
shows that the function $F(z;k)$ of the form (\ref{2.2.4}) is 
transformed as
	\begin{equation}
F(z;k) \rightarrow \left[ \frac{ \exp(-b^{\ast}z/a^{\ast}) 
}{ (a^{\ast})^{2k} } \right] \sum_{n=0}^{\infty} 
\frac{ C_n }{ \sqrt{n!\Gamma(n+2k)} } R_{n}(z)
	\end{equation}
where the function $R_{n}(z)$ can be written by means of the Laguerre
polynomial:
	\begin{equation}
R_{n}(z) = (b/a^{\ast})^n n! L_{n}^{2k-1}(z/a^{\ast}b) .
	\end{equation}
We see that the function $F(z;k)$ is multiplied by the factor in the
square brackets and $z^n$ is replaced by $R_{n}(z)$.

We now apply the above transform to an example that will be used
later. We consider the general eigenvalue equation 
	\begin{equation}
( \beta_{1} K_{1} + \beta_{2} K_{2} +\beta_{3} K_{3} ) 
|\lambda,k\rangle = \lambda |\lambda,k\rangle .  \label{3.5}
	\end{equation}
Here $\beta_{i}$ ($i=1,2,3$) are arbitrary complex numbers. In the
analytic representation in the unit disk this eigenvalue equation 
becomes the first-order linear homogeneous differential equation:
	\begin{equation}
(\beta_{+} + \beta_{3}\zeta + \beta_{-}\zeta^{2}) 
\frac{ {\rm d} G(\zeta;\lambda,k)}{ {\rm d} \zeta} + 
(2k\beta_{-}\zeta + k\beta_{3}-\lambda) G(\zeta;\lambda,k) = 0          
\label{3.6}
	\end{equation}
where we have defined
$\beta_{\pm} \equiv \frac{1}{2} (\beta_{1} \pm {\rm i} \beta_{2})$.
Equation (\ref{3.6}) can be rewritten as
	\begin{equation}
(\beta_{+}\rho^{2} + \beta_{3}\rho + \beta_{-}) 
\frac{ {\rm d} G(1/\rho;\lambda,k)}{ {\rm d} \rho} - 
\left( \frac{2k\beta_{-}}{\rho} + k\beta_{3} -\lambda\right) 
G(1/\rho;\lambda,k) = 0 .  \label{3.7}
	\end{equation}
In the Barut-Girardello analytic representation the eigenvalue equation 
(\ref{3.5}) becomes the second-order linear homogeneous differential 
equation with linear coefficients:
	\begin{equation}
\beta_{+}z \frac{ {\rm d}^{2} F(z;\lambda,k)}{ {\rm d} z^{2}} + 
(\beta_{3}z + 2k\beta_{+}) \frac{ {\rm d} F(z;\lambda,k)}{ 
{\rm d} z} + (\beta_{-}z + k\beta_{3} - \lambda) F(z;\lambda,k) = 0 .   
\label{3.8}
	\end{equation} 
It is easy to check that by substituting the Laplace transform 
(\ref{3.3}) into equation (\ref{3.7}), we obtain equation 
(\ref{3.8}); conversely, substituting the inverse Laplace 
transform (\ref{3.4}) into equation (\ref{3.8}), we obtain 
equation (\ref{3.7}).
It is interesting to note that the
normalization condition for a solution of equation (\ref{3.8}) is 
equivalent to the analyticity condition for a solution of equation 
(\ref{3.7}). This condition determines the allowed values of $\lambda$.
The technique of analytic representations allows us to obtain the 
various states associated with the SU(1,1) Lie group and to derive 
analytic expressions for expectation values over these states. 

\section{A ``weak'' resolution of the identity in terms of
SU(1,1) coherent states}

\noindent
Resolutions of the identity in terms of a certain set of states
are very important because they allow the practical use of these
states as a basis in the Hilbert space. There are situations where
it can be proved that a certain set of states is overcomplete but no
resolution of the identity is known. There is a need in these cases to
develop weaker concepts than the usual resolutions of the identity,
in order to be able to use these states as a basis. The concept of 
frames used widely in the subject of wavelets is such an example. 
In the case of the SU(1,1) Perelomov coherent states the resolution
of the identity (\ref{2.1.2}) is valid for $k>1/2$ only.
In this section we present another resolution of the identity which is 
valid for both $k>1/2$ and $k<1/2$.

We substitute $ {\rm d}^{2} \zeta = \frac{1}{2} {\rm d} t {\rm d} \phi$, 
(where $\zeta = \sqrt{t}\exp( {\rm i} \phi)$) in (\ref{2.1.2}) and 
integrate over the angle $\phi$, to get
	\begin{equation}
I = \sum_{n=0}^{\infty} \frac{\Gamma(n+2k)}{\Gamma(n+1)\Gamma(2k-1)}
\left[ \int_{0}^{1} t^{n} (1-t)^{2k-2} {\rm d} t \right] 
|n,k \rangle\langle n,k|.   \label{4.1}
	\end{equation} 
For $k>1/2$, the integral over $t$ is the Euler 
Beta function. Then one obtains the desired result 
$I = \sum_{n=0}^{\infty} |n,k \rangle\langle n,k|$. However, for 
$k<1/2$ the integral in (\ref{4.1}) is divergent.

We consider the following relation for the 
Beta function (Erd\'{e}lyi {\em et al} 1953a, sec 1.6):
	\begin{eqnarray}
& & {\rm B}(x,y) = \frac{1}{2 {\rm i} \sin(\pi y)} \oint_{{\cal C}} 
t^{x-1} (t-1)^{y-1} {\rm d} t  \nonumber \\
& & {\rm Re}\, x > 0  \;\;\;\;\;\;\;\; 
|{\rm arg}\,(t-1)| \leq \pi  \;\;\;\;\;\;\;\; 
y \neq 0, \pm 1, \pm 2, \ldots . \label{4.5}	
	\end{eqnarray}
The contour ${\cal C}$ is a single loop that goes from the origin 
up to one below the real axis, turns back around the point $t=1$ in 
the counter-clockwise direction, and goes back above the real axis
up to zero. 

We next point out that in equation (\ref{2.1.3})
many functions converge in a disk that is larger than the unit disk.
We call $H({\cal D}(1+\epsilon))$ the subspace of the Hardy space
that contains all the functions that converge in the disk
${\cal D}(1+\epsilon) = \{|\zeta|< 1+\epsilon \}$ (where $\epsilon > 0$).
Clearly if $\epsilon_1 > \epsilon_2$ then the
$H({\cal D}(1+\epsilon_1))$ is a subspace of $H({\cal D}(1+\epsilon_2))$.
As $\epsilon$ goes to 0 (from above),
the $H({\cal D}(1+\epsilon))$ becomes the Hardy space.
Using (\ref{4.5}), we can prove that for any positive $k$ apart from
integers and half-integers and for any two states in
$H({\cal D}(1+\epsilon))$ (where $\epsilon$ is any positive number), 
the scalar product can be written in the form 
        \begin{equation}
\langle \Psi_1 | \Psi_2 \rangle =
\frac{-(2k-1)\exp(2\pi {\rm i} k)}{4\pi {\rm i} \sin(2\pi k)}\,
\oint_{{\cal C}}  \frac{ {\rm d} t}{(1-t)^{2-2k}} \int_{0}^{2\pi} 
{\rm d} \phi \, [G_{1}(\zeta;k)]^{\ast} G_{2}(\zeta;k)  .  \label{4.8}
        \end{equation}
The contour ${\cal C}$ goes around 1 but is entirely within 
${\cal D}(1+\epsilon)$. In order to prove this result, we substitute
the functions $G(\zeta;k)$ of the form (\ref{2.1.3}) into (\ref{4.8}).
Taking $\zeta = \sqrt{t} \exp( {\rm i} \phi)$, we integrate over the
angle $\phi$. Then equation (\ref{4.8}) reads
        \begin{equation}
\langle \Psi_1 | \Psi_2 \rangle = \sum_{n=0}^{\infty} 
C_{1 n}^{\ast} C_{2 n} \frac{\Gamma(n+2k)}{n! \Gamma(2k)}
\frac{(2k-1)}{2 {\rm i} \sin(2\pi k -\pi)}\,
\oint_{{\cal C}} t^{n} (t-1)^{2k-2} {\rm d} t .  \label{4.8a}
        \end{equation}
We can take the summation out of the integral because the 
two states belong to $H({\cal D}(1+\epsilon))$, i.e. their analytic 
functions are convergent on the contour ${\cal C}$.
We see that the integral in (\ref{4.8a})
is exactly of the form (\ref{4.5}) with $x=n+1$, $y=2k-1$. Therefore 
formula (\ref{4.5}) can be used for any positive $k$ apart from
integers and half-integers. Using the result (\ref{4.5}), we rewrite
equation (\ref{4.8a}) in the form
        \begin{equation}
\langle \Psi_1 | \Psi_2 \rangle = \sum_{n=0}^{\infty} 
C_{1 n}^{\ast} C_{2 n} \frac{\Gamma(n+2k)}{n! \Gamma(2k)}
(2k-1) {\rm B}(n+1,2k-1) .  \label{4.8b}
        \end{equation}
If we express the Beta function in terms of the Gamma functions,
equation (\ref{4.8b}) reduces to the usual form of the scalar product:
$\langle \Psi_1 | \Psi_2 \rangle = \sum_{n=0}^{\infty} 
C_{1 n}^{\ast} C_{2 n}$.

Equation (\ref{4.8}) is a ``weak'' resolution 
of the identity which we express as 
	\begin{equation}
I = \frac{-(2k-1)\exp(2\pi {\rm i} k)}{4\pi {\rm i} \sin(2\pi k)}
\oint_{{\cal C}} \frac{ {\rm d} t}{(1-t)^{2}} \int_{0}^{2\pi} {\rm d} 
\phi \, |\zeta,k \rangle\langle \zeta,k|.   \label{4.6}
	\end{equation} 
Although the contour ${\cal C}$ goes outside the unit disk, where the 
SU(1,1) coherent states are not normalisable, this equation has to be 
understood in a weak sense in conjuction with equation (\ref{4.8}).
A completeness relation for SU(1,1) coherent states
is also discussed by W\"{u}nsche (1992) using a rigged Hilbert space.

\section{Two-photon realization of SU(1,1) in quantum optics}

\subsection{Resolution of the identity}

\noindent
We consider the two-photon realization of the SU(1,1) 
Lie algebra:
	\begin{equation}
K_{+} = \frac{1}{2} a^{\dagger 2}  \;\;\;\;\;\;\;\;\;
K_{-} = \frac{1}{2} a^{2}  \;\;\;\;\;\;\;\;\;
K_{3} = \frac{1}{4} (aa^{\dagger}+a^{\dagger}a)  \label{5.0}
	\end{equation}
where $a$ and $a^{\dagger}$ are the boson annihilation and creation
operators that satisfy the canonical commutation relation
$[a,a^{\dagger}]=I$. The Casimir operator is $K^{2} = -3/16$.
Therefore, there are two irreducible representations: $k=1/4$
and $k=3/4$ (Bishop and Vourdas 1987). The state space 
${\cal H}_{1/4}$ is the even Fock
subspace with the orthonormal basis consisting of even number 
eigenstates $|n,\frac{1}{4}\rangle = |2n\rangle$ ($n=0,1,2,\ldots$); 
the state space ${\cal H}_{3/4}$ is the odd Fock subspace with the 
orthonormal basis consisting of odd number eigenstates 
$|n,\frac{3}{4}\rangle = |2n+1\rangle$ ($n=0,1,2,\ldots$). 
A state $|\Psi\rangle$ in the total Fock space can be written as
	\begin{equation}
\mbox{\hspace{-0.8cm}}
|\Psi\rangle = |\Psi ; e \rangle + |\Psi ; o \rangle = 
\sum_{n=0}^\infty  C_{n}^{(e)} |n ;\mbox{$\frac{1}{4}$} \rangle 
+ \sum_{n=0}^\infty C_n ^{(o)} |n ;\mbox{$\frac{3}{4}$} \rangle 
=  \sum_{n=0}^\infty (C_{2n} |2n \rangle + C_{2n+1} |2n + 1 \rangle)  
\label{state}
	\end{equation}
with the normalisation
	\begin{equation}
\sum_{n=0}^\infty |C_{n}^{(e)}|^2 = {\cal N}_e \;\;\;\;\;\;\;\;\;\;
\sum_{n=0}^\infty |C_{n}^{(o)}|^2 = {\cal N}_o \;\;\;\;\;\;\;\;\;\;
{\cal N}_e + {\cal N}_o = 1 .
	\end{equation}
Here $e$ and $o$ indicate even ($k=1/4$) and odd ($k=3/4$) subspaces, 
correspondingly. It is clear that $C_{2n} = C_{n}^{(e)}$, 
$C_{2n+1} = C_{n}^{(o)}$.

The unitary group operator $\Omega(\xi) \in$ SU(1,1)/U(1) for the
two-photon realization is the well-known squeezing operator 
$S(\xi)$ (Stoler 1970, 1971, Yuen 1976):
	\begin{equation}
S(\xi) = \exp\left( \xi K_{+} - \xi^{\ast} K_{-} \right) =
\exp\left( \frac{\xi}{2} a^{\dagger 2}  - \frac{\xi^{\ast}}{2} a^{2} 
\right) .     \label{5.1}
	\end{equation}
Therefore, the SU(1,1) coherent states are the squeezed states. 
For $k=1/4$ one gets the squeezed vacuum,
	\begin{equation}
| \zeta,\mbox{$\frac{1}{4}$} \rangle = S(\xi) 
|0\rangle = (1-|\zeta|^{2})^{1/4} \sum_{n=0}^{\infty} 
\frac{ \sqrt{(2n)!} }{ 2^{n} n!} \zeta^{n} |2n\rangle  \label{5.2}
	\end{equation}
while for $k=3/4$ one gets the squeezed ``one photon'' state,
	\begin{equation}
| \zeta,\mbox{$\frac{3}{4}$} \rangle = S(\xi) 
|1\rangle = (1-|\zeta|^{2})^{3/4} \sum_{n=0}^{\infty} \frac{ 
\sqrt{(2n+1)!}}{2^{n} n!} \zeta^{n} |2n+1\rangle .    \label{5.3}
	\end{equation}
As before, $\zeta = (\xi/|\xi|) \tanh |\xi|$. It is also possible to
define the parity-dependent squeezing operator (Brif {\em et al} 
1996) that imposes different squeezing transformations on the even
and odd subspaces of the Fock space.

Using the results of the previous section, we obtain the 
following resolutions of the identity (in a weak sense, 
as explained above):
	\begin{eqnarray}
\frac{1}{8\pi} \oint_{{\cal C}} \frac{ {\rm d} t}{(1-t)^{2}} 
\int_{0}^{2\pi} {\rm d} \phi \, | \zeta,\mbox{$\frac{1}{4}$}
\rangle \langle \zeta,\mbox{$\frac{1}{4}$} | 
& = & \sum_{n=0}^{\infty} |2n \rangle\langle 2n|    \label{5.4} \\
-\frac{1}{8\pi} \oint_{{\cal C}} \frac{ {\rm d} t}{(1-t)^{2}} 
\int_{0}^{2\pi} {\rm d} \phi \, | \zeta,\mbox{$\frac{3}{4}$} 
\rangle \langle \zeta,\mbox{$\frac{3}{4}$} | 
& = & \sum_{n=0}^{\infty} |2n+1 \rangle\langle 2n+1|    \label{5.5}
	\end{eqnarray}
where $\zeta = \sqrt{t}\, {\rm e}^{ {\rm i} \phi}$.

\subsection{A synthesis of the $k=1/4$ and $k=3/4$ 
Barut-Girardello representations and its relation to the Bargmann
representation}

In the two-photon realization the Barut-Girardello 
eigenvalue equation takes the form
	\begin{equation}
a^{2} |z,k\rangle = 2z |z,k\rangle .    \label{5.6}
	\end{equation}
Therefore, it is not difficult to see that the Barut-Girardello states 
$|z,k\rangle$ coincide with the even and odd coherent states,
	\begin{eqnarray}
|z,\mbox{$\frac{1}{4}$} \rangle & = & |\alpha\rangle_{e} 
= \frac{1}{ \sqrt{2\left(1+{\rm e}^{-2|\alpha|^{2}}\right)} } 
\left( |\alpha\rangle + |-\alpha\rangle \right)   \label{5.7} \\
|z,\mbox{$\frac{3}{4}$} \rangle & = & |\alpha\rangle_{o} 
= \frac{1}{ \sqrt{2\left(1-{\rm e}^{-2|\alpha|^{2}}\right)} } 
\left( |\alpha\rangle - |-\alpha\rangle \right)   \label{5.8}
	\end{eqnarray}
for $k=1/4$ and $k=3/4$, respectively. Here $|\alpha\rangle$ 
are the familiar Glauber coherent states (Glauber 1963).
The even and odd coherent states were introduced by Dodonov 
{\em et al} (1974) and they satisfy the eigenvalue equation
	\begin{equation}
a^{2} |\alpha\rangle_{e,o} = \alpha^{2} |\alpha\rangle_{e,o}  .    
\label{5.10}
	\end{equation}
The above comments indicate that there must be a relation
between a synthesis of the $k=1/4$ and $k=3/4$ 
Barut-Girardello representations and the Bargmann representation in 
the $\alpha$ plane (Bargmann 1961), with $\alpha^{2} = 2z$. 

In order to demonstrate this explicitly we consider the $k=1/4$ 
and $k=3/4$ Barut-Girardello representations for the even and odd 
components, correspondingly, of the state $|\Psi\rangle$ of equation
(\ref{state}) with $z=\alpha^{2}/2$:
\begin{equation}
F(\mbox{$\frac{1}{2}$}\alpha^2 ; \mbox{$\frac{1}{4}$}) = 
\frac{1}{ \sqrt{{\cal N}_{e}} } \sum_{n = 0}^{\infty}
\frac{ C_{n}^{(e)} }{ \sqrt{n!\Gamma(n+\frac{1}{2})} } 
\left( \frac{\alpha^2}{2} \right)^n 
= \frac{ \pi^{-1/4} }{ \sqrt{{\cal N}_{e}} } \sum_{n = 0}^{\infty} 
\frac{ C_{2n} }{\sqrt{(2n)!}}\, \alpha^{2n}
\end{equation}
\begin{equation}
\mbox{\hspace{-0.8cm}}
F(\mbox{$\frac{1}{2}$}\alpha^2 ; \mbox{$\frac{3}{4}$}) = 
\frac{1}{ \sqrt{{\cal N}_{o}} } \sum_{n = 0}^{\infty}
\frac{ C_{n}^{(o)} }{ \sqrt{n!\Gamma(n+\frac{3}{2})} } 
\left( \frac{\alpha^2}{2} \right)^n
=  \frac{ \pi^{-1/4} }{ \sqrt{{\cal N}_{o}} }  
\left( \frac{\alpha}{\sqrt{2}} \right)^{-1} \sum_{n = 0}^{\infty} 
\frac{ C_{2n+1} }{\sqrt{(2n+1)!}}\, \alpha^{2n+1}
\end{equation}
It is clear that
\begin{equation}
B(\alpha) = \pi^{1/4} \left[ \sqrt{{\cal N}_{e}}
F(\mbox{$\frac{1}{2}$}\alpha^2 ; \mbox{$\frac{1}{4}$}) + 
\sqrt{{\cal N}_{o}} \frac{\alpha}{\sqrt{2}} 
F(\mbox{$\frac{1}{2}$}\alpha^2 ; \mbox{$\frac{3}{4}$}) \right] 
\end{equation}
is the Bargmann representation (Bargmann 1961) 
for the state $|\Psi\rangle$. This relation between the Bargmann
and Barut-Girardello analytic representations has been discussed
by Basu (1992). This is a particular case of a more general relation
between the Bargmann representation for the para-Bose oscillator 
characterized by the parameter $h_0 = 2k$ and the SU(1,1) 
Barut-Girardello representations with the indices $k$ and 
$k+\frac{1}{2}$ (Sharma {\em et al} 1978, Mukunda {\em et al} 1980, 
Sharma {\em et al} 1981).
It is also well known that in the Bargmann representation 
$a^{\dagger} = \alpha $, $a = {\rm d} / {\rm d} \alpha$,
and consequently the SU(1,1) generators are the operators of the form
	\begin{equation}
K_{+} = \frac{\alpha^{2}}{2}  \;\;\;\;\;\;\;\;\;
K_{-} = \frac{1}{2} \frac{ {\rm d}^{2} }{ {\rm d} \alpha^{2}} 
\;\;\;\;\;\;\;\;\;
K_{3} = \frac{\alpha}{2}\frac{ {\rm d} }{ {\rm d} \alpha} + \frac{1}{4}  
\label{5.14}
	\end{equation}
which is the same as equation (\ref{2.2.6}) for the Barut-Girardello
representation with $k=1/4$ and $z=\alpha^{2}/2$.
  
As an application we consider equation 
(\ref{3.5}) which in the Bargmann representation is a second-order 
differential equation that has two independent solutions: the first
corresponds to $k=1/4$, and the second to $k=3/4$.
We illustrate this by considering solutions of
equation (\ref{3.8}). For $\beta_{+} \neq 0$, 
$\beta_{3}^{2} - \beta_{1}^{2} - \beta_{2}^{2} \neq 0$, two 
independent solutions are:
	\begin{eqnarray}
& & F_{e}(z=\alpha^{2}/2;\lambda,k) = \exp\left( 
\frac{\Delta-\beta_{3}}{\beta_{1}+ {\rm i} \beta_{2}} 
\frac{\alpha^{2}}{2} \right) \,
\Phi\left( k-\frac{\lambda}{\Delta} ; 2k ; \frac{-\Delta}{
\beta_{1}+ {\rm i} \beta_{2}} \alpha^{2} \right)   \label{5.15} \\
& & F_{o}(z=\alpha^{2}/2;\lambda,k) = \exp\left( 
\frac{\Delta-\beta_{3}}{\beta_{1}+ {\rm i} \beta_{2}} 
\frac{\alpha^{2}}{2} \right)  \alpha^{2-4k} \,
\Phi\left( \tilde{k}-\frac{\lambda}{\Delta} ; 2\tilde{k} ; 
\frac{-\Delta}{\beta_{1}+ {\rm i} \beta_{2}} \alpha^{2} \right)   
\label{5.16}
	\end{eqnarray}
where $\Delta \equiv \sqrt{\beta_{3}^{2} - \beta_{1}^{2} - 
\beta_{2}^{2}}$, $\tilde{k} \equiv 1-k$, and $\Phi(a;b;x)$ is
the confluent hypergeometric function (the Kummer function).
The two solutions differ by the factor $\alpha^{2-4k}$ and by the
replacement $k \rightarrow \tilde{k}$. When $k=1/4$, we have
$\tilde{k} = 3/4$ and $\alpha^{2-4k} = \alpha$. Thus, if the even
solution $F_{e}$ corresponds to $k=1/4$, the other solution is
odd and it corresponds to $\tilde{k} = 3/4$.

\subsection{Analytic representations based on displaced
squeezed states}

\noindent
In the preceding section we have shown how an analytic representation
can be constructed by using the squeezed vacuum states. However, in
quantum optics one usually meets squeezing of the Glauber coherent 
states. The corresponding states are referred to as the displaced 
squeezed states:
	\begin{equation}
|\zeta,\eta\rangle = D(\eta) S(\xi) |0\rangle .    \label{6.2}
	\end{equation}
Here $S(\xi)$ is the squeezing operator of equation (\ref{5.1})
and $D(\eta) = \exp(\eta a^{\dagger} - \eta^{\ast} a)$ is the
displacement operator that generates the Glauber coherent states:
$|\alpha\rangle = D(\alpha) |0\rangle$. We would like to construct
the analytic representation based on the displaced squeezed states.
Therefore we use the displaced two-photon realization of the SU(1,1) 
Lie algebra:
	\begin{eqnarray}
& & K_{+}(\eta) = D(\eta) K_{+} D^{-1}(\eta) = \mbox{$\frac{1}{2}$}
(a^{\dagger} - \eta^{\ast})^{2}  \label{6.3} \\
& & K_{-}(\eta) = D(\eta) K_{-} D^{-1}(\eta) = \mbox{$\frac{1}{2}$}
(a - \eta)^{2}  \label{6.4} \\
& & K_{3}(\eta) = D(\eta) K_{3} D^{-1}(\eta) = \mbox{$\frac{1}{2}$}
(a^{\dagger} - \eta^{\ast}) (a-\eta) + \mbox{$\frac{1}{4}$} .  
\label{6.5} 
	\end{eqnarray}
Since the above operators are produced by a similarity transformation,
commutation relations and the Casimir operator remain unchanged.
The displaced SU(1,1) generators can produce both squeezing and 
displacing transformations. The orthonormal basis corresponding to
the displaced two-photon realization of SU(1,1) consists of the
displaced Fock states 
(de Oliveira {\em et al\/} 1990, W\"{u}nsche 1991)
	\begin{equation}
|n,k\rangle_{D} = D(\eta) |n,k\rangle = \left\{ \begin{array}{l}
D(\eta) |2n\rangle \;\;\;\;\;\;\; k=1/4 \\
D(\eta) |2n+1\rangle \;\;\;\;\;\;\; k=3/4 .  \end{array}
\right.                     \label{6.6}
	\end{equation}
The SU(1,1) coherent states are obtained by the action of the unitary 
operator
	\begin{equation}
S_{D}(\xi) = D(\eta) S(\xi) D^{-1}(\eta) = \exp [ \xi K_{+}(\eta)
- \xi^{\ast} K_{-}(\eta)]       \label{6.7}
	\end{equation}
on the lowest state $|0,k\rangle_{D} = D(\eta) |0,k\rangle$. As result,
we obtain the displaced squeezed states
	\begin{equation}
|\zeta,k\rangle_{D} = D(\eta) S(\xi) |0,k\rangle = D(\eta) 
|\zeta,k\rangle .      \label{6.8}
	\end{equation}
These are the $|\zeta,\eta\rangle$ states of equation (\ref{6.2})
when $k=1/4$; for $k=3/4$ the SU(1,1) coherent states are 
$|1,\zeta,\eta\rangle = D(\eta) S(\xi) |1\rangle$. Apart from the
well-known resolution of the identity 
	\begin{equation}
I = \frac{1}{\pi} \int {\rm d}^{2} \eta |\zeta,\eta\rangle \langle \zeta,
\eta|       \label{6.9}
	\end{equation}
we have now another resolution of the identity for the squeezed states:
	\begin{equation}
I = \frac{1}{8\pi} \oint_{{\cal C}} \frac{ {\rm d} t }{ (1-t)^{2} }
\int_{0}^{2\pi} {\rm d} \phi \left( |\zeta,\eta\rangle \langle \zeta,\eta|
- |1,\zeta,\eta\rangle \langle 1,\zeta,\eta| \right) .  \label{6.10}
	\end{equation}
This result follows immediately from relations (\ref{5.4}) and 
(\ref{5.5}). Then a state $|\Psi\rangle$ can be represented as 
	\begin{equation}
\mbox{\hspace{-0.8cm}}
|\Psi\rangle = \frac{1}{8\pi} \oint_{{\cal C}} \frac{ {\rm d} t 
}{ (1-t)^{2} } \int_{0}^{2\pi} {\rm d} \phi \left[ (1-t)^{1/4} 
G_{D}(\zeta^{\ast};\mbox{$\frac{1}{4}$}) |\zeta,\eta\rangle - 
(1-t)^{3/4} G_{D}(\zeta^{\ast};\mbox{$\frac{3}{4}$})
|1,\zeta,\eta\rangle \right]   \label{6.11}
	\end{equation}
where 
	\begin{equation}
G_{D}(\zeta;k) = (1-|\zeta|^{2})^{-k} {}_{D}\langle \zeta^{\ast},k|
\Psi\rangle   .       \label{6.12}
	\end{equation}

We can use this ``displaced'' version of the SU(1,1) coherent-state 
analytic representation for analyzing the spectrum of the
squeezed and displaced harmonic oscillator (Zhang {\em et al} 1990)
	\begin{equation}
H = \omega \left( a^{\dagger}a + \mbox{$\frac{1}{2}$} \right) 
+ \frac{g}{2} a^{\dagger 2} + \frac{g^{\ast}}{2} a^{2} 
+ f a^{\dagger} + f^{\ast} a  .          \label{6.1}
	\end{equation}
Here $\omega$ is a real positive frequency and $g$, $f$ are
arbitrary complex parameters. The Hamiltonian (\ref{6.1}) is a linear 
combination of generators of the maximal symmetry group for the 
quantum harmonic oscillator (Niederer 1973). This group is the 
semidirect product of the SU(1,1) group and the Weyl-Heisenberg 
group (Weyl 1950) whose generators are $a$, $a^{\dagger}$ and $I$.
It is well known (Yuen 1976, Beckers and Debergh 1989, Zhang {\em et al} 
1990) that a quantum state evolved by the Hamiltonian of type (\ref{6.1}) 
is displaced and squeezed. The eigenstates and eigenvalues
of the Hamiltonian (\ref{6.1}) were extensively studied by Lo (1991b), 
Nagel (1995) and W\"{u}nsche (1995).
Eigenvalue problems for Hermitian combinations of SU(1,1) generators 
were first considered by Lindblad and Nagel (1970) and Solomon (1971). 
We would like to approach this problem by using the squeezed-state 
analytic representation.

Using the displaced SU(1,1) generators (\ref{6.3})-(\ref{6.5}),
the Schr\"{o}dinger equation $H|E\rangle = E|E\rangle$ for the
Hamiltonian (\ref{6.1}) can be written in the form
	\begin{equation}
[ 2\omega K_{3}(\eta) + g K_{+}(\eta) + g^{\ast} K_{-}(\eta) ]
|E\rangle = (E+\delta) |E\rangle     \label{6.16}
	\end{equation}
	\begin{equation}
\eta = \frac{ g f^{\ast} - \omega f }{ \omega^{2} - |g|^{2} }
\;\;\;\;\;\;\;\;\;\;\;\;
\delta = \frac{ \omega |f|^{2} - {\rm Re}\, 
(g f^{\ast 2}) }{ \omega^{2} - |g|^{2} }  .   \label{6.15}
	\end{equation}
Using the squeezed-state analytic representation, we write 
(\ref{6.16}) as a first-order linear differential equation of 
type (\ref{3.6}):
	\begin{equation}
(g^{\ast} + 2\omega\zeta + g\zeta^{2}) \frac{ {\rm d} G_{D}(\zeta;E,k) 
}{ {\rm d} \zeta } + (2kg\zeta + 2k\omega -E -\delta) G_{D}(\zeta;E,k)
= 0 \label{6.17}
	\end{equation}
where the analytic function $G_{D}(\zeta;E,k)$ is defined by 
(\ref{6.12}). The solution of equation (\ref{6.17}) is 
	\begin{equation}
G_{D}(\zeta;E,k) = G_{0} (\zeta + \chi_{-})^{-k+u}
(\zeta + \chi_{+})^{-k-u}       \label{6.18}
	\end{equation}
where $G_{0}$ is a normalization factor and we have defined 
	\begin{equation}
\chi_{\pm} \equiv ( \omega \pm \Delta )/g  
\;\;\;\;\;\;\;\;\;\;\;\;
u \equiv ( E+\delta )/( 2\Delta ) \;\;\;\;\;\;\;\;\;\;\;\;
\Delta \equiv \sqrt{\omega^{2} - |g|^{2}} .
\label{6.20}
	\end{equation}
Here we should distinguish between the two possibilities: $\omega < |g|$
and $\omega > |g|$. When $\omega < |g|$, then $|\chi_{+}| = |\chi_{-}|
= 1$ and the system has only a continuous spectrum (Lo 1990, 1991b). 
We consider the case of $\omega > |g|$ in which 
$\chi_{+} = 1/\chi_{-}^{\ast}$ and the system has a discrete spectrum. 
In this case $|\chi_{-}| < 1$ and the analyticity condition requires 
	\begin{equation}
-k+u = l =0,1,2,\ldots  \label{6.21}
	\end{equation}
that leads to the quantization condition
	\begin{equation}
E = E_{l}(k) = 2\sqrt{\omega^{2}-|g|^{2}}\, (k+l) 
- \frac{ \omega |f|^{2} - {\rm Re}\, (g f^{\ast 2}) }{ 
\omega^{2} - |g|^{2} } .           \label{6.22}                    
	\end{equation}
Then the function $G_{D}(\zeta;E,k)$ takes the form (with 
$\chi \equiv \chi_{-} = 1/\chi_{+}^{\ast}$)
	\begin{equation}
G_{D}(\zeta;l,k) = G_{0} (\zeta+\chi)^{l} 
(\zeta + 1/\chi^{\ast})^{-2k-l}    .    \label{6.23}
	\end{equation}
It is clear that for any value of $(\omega - |g|) > 0$ there 
exists  $\epsilon > 0$ such that this function belongs to the subspace 
$H({\cal D}(1+\epsilon))$.
The analyticity condition (\ref{6.21}) cannot be 
simultaneously satisfied for $k=1/4$ and $k=3/4$. Therefore, if
$G_{D}(\zeta;l,\mbox{$\frac{1}{4}$})$ is analytic,
$G_{D}(\zeta;l,\mbox{$\frac{3}{4}$})$ must be zero
(the trivial solution), and vice versa. Thus for $\omega > |g|$ 
the Hamiltonian (\ref{6.1}) has two distinct series of eigenstates 
and eigenvalues: one for $k=1/4$ and the other for $k=3/4$. The 
eigenstates depend on the Bargmann index $k$ and therefore they 
belong to only one of the SU(1,1) irreducible representations. 

It can be seen that the function (\ref{6.23}) represents 
the squeezed and displaced Fock states (Kral 1990, Lo 1991a, b):
	\begin{equation}
|n,\zeta_{0},\eta\rangle = D(\eta) S(\xi_{0}) |n\rangle .         
\label{6.24}
	\end{equation}
where $\xi_{0} = (s/2)\, {\rm e}^{ {\rm i} \theta}$, $\eta = 
r\, {\rm e}^{ {\rm i} \vartheta}$ are related to $\omega$, $g$, $f$ by 
	\begin{equation}
\omega/\Delta = \cosh s \;\;\;\;\;\;\;\; 
g/\Delta = - \sinh s\, {\rm e}^{ {\rm i} \theta} \;\;\;\;\;\;\;\;
f/\Delta = \eta \left( \sinh s\, {\rm e}^{ {\rm i} (\theta - 2\vartheta)} 
- \cosh s \right)    \label{6.28}
	\end{equation}
and the integer $n$ is given by
	\begin{equation}
n = 2l + 2k - \frac{1}{2} = \left\{ \begin{array}{l}
2l \;\;\;\;\;\;\; k=1/4 \\
2l+1 \;\;\;\;\;\;\; k=3/4 \end{array} \right.   .  \label{6.29}
	\end{equation}
Thus the energy eigenstates corresponding to $k=1/4$ ($3/4$) are the
squeezed and displaced even (odd) Fock states.

\section{Conclusions}

Analytic representations exploit the powerful theory of analytic 
functions in a quantum mechanical context. In this paper we have 
studied various aspects of the analytic representation in the
unit disk and the Barut-Girardello representation. We have shown 
that the two are related through a Laplace transform. We have also 
considered the resolution of the identity in terms of the 
SU(1,1) Perelomov coherent states which is known to be valid
only for $k>1/2$. With an analytic continuation we have derived
a ``weak'' resolution of the identity which is valid even in the region
$k<1/2$.

All these ideas have been applied in the context of squeezed
states in quantum optics. We have shown how a synthesis of the $1/4$
and $3/4$ representations is related to the Bargmann representation.
We have also considered analytic representations based on displaced
squeezed states, and used them for the study of the displaced and 
squeezed harmonic oscillator. The results demonstrate that apart from 
their theoretical merit, analytic reprentations can also be useful 
in practical calculations.

\section*{Acknowledgements}

CB thanks Professor M S Marinov for valuable discussions and
gratefully acknowledges the financial help from the Technion.
AV thanks Professor A~W\"{u}nsche for helpful discussions on SU(1,1)
coherent states and gratefully acknowledges support from the British 
council in the form of a travel grant. 
AM was supported by the Fund for Promotion of Research
at the Technion, by the Technion -- VPR Fund, and by GIF --- 
German-Israeli Foundation for Research and Development.

%\newpage

\section*{References}

\begin{description}
\item Agarwal G S 1988 {\em J. Opt. Soc. Am.} B {\bf 5} 1940
\item Bar-Moshe D and Marinov M S 1994 {\em J. Phys. A: Math. Gen.}
      {\bf 27} 6287
\item Bargmann V 1947 {\em Ann. Math.} {\bf 48} 568
\item ------ 1961 {\em Commun. Pure Appl. Math.} {\bf 14} 187
\item Barut A O and Girardello L 1971 {\em Commun. Math. Phys.} 
      {\bf 21} 41
\item Basu D 1992 {\em J. Math. Phys.} {\bf 33} 114
\item Beckers J and Debergh N 1989 {\em J. Math. Phys.} 
      {\bf 30} 1732
\item Berezin F A 1974 {\em Sov. Math. Izv.} {\bf 38} 1116
\item ------ 1975a {\em Sov. Math. Izv.} {\bf 39} 363
\item ------ 1975b {\em Commun. Math. Phys.} {\bf 40} 153
\item Bishop R F and Vourdas A 1987 {\em J. Phys. A: Math. Gen.} 
      {\bf 20} 3727
\item Brif C 1995 {\em Quantum Semiclas. Opt.} {\bf 7} 803
\item Brif C and Ben-Aryeh Y 1994a {\em Quantum Opt.} {\bf 6} 391 
\item ------ 1994b {\em J. Phys. A: Math. Gen.} {\bf 27} 8185 
\item ------ 1996 {\em Quantum Semiclas. Opt.} {\bf 8} 1
\item Brif C, Mann A and Vourdas A 1996 {\em J. Phys. A: 
      Math. Gen.} {bf 29} 2053
\item Bu\v{z}ek V 1990 {\em J. Mod. Opt.} {\bf 37} 303
\item de Oliveira F A M, Kim M S, Knight P L and Bu\v{z}ek V 1990
      {\em Phys. Rev.} A {\bf 41} 2645
\item Dodonov V V, Malkin I A and Man'ko V I 1974 {\em Physica} 
      {\bf 72} 597
\item Erd\'{e}lyi {\em et al} (ed) 1953a {\em 
      Bateman Manuscript Project: Higher Transcendental Functions} 
      (New York: McGraw-Hill)
\item ------ 1953b {\em Bateman Manuscript Project: 
      Tables of Integral Transforms} (New York: McGraw-Hill)
\item Gerry C C 1983 {\em J. Phys. A: Math. Gen.} {\bf 16} L1
\item Glauber R J 1963 {\em Phys. Rev.} A {\bf 131} 2766
\item Fock V A 1928 {\em Z. Phys.} {\bf 49} 339
\item Kim Y S and Noz M E 1986 {\em Theory and Applications of the
      Poincare Group} (Dordrecht: Reidel)
\item Kuriyan J G, Mukunda N and Sudarshan E C G 1968 {\em J. Math.
      Phys.} {\bf 9} 2100
\item Kral P 1990 {\em J. Mod. Opt.} {\bf 37} 889
\item Lindblad G and Nagel B 1970 {\em Ann. Inst. Henri Poinc.} A 
      {\bf 13} 27
\item Lo C F 1990 {\em Phys. Rev.} A {\bf 42} 6752
\item ------ 1991a {\em Phys. Rev.} A {\bf 43} 404
\item ------ 1991b {\em Quantum Opt.} {\bf 3} 333
\item Mukunda N, Sudarshan E C G, Sharma J K and Mehta C L 1980
      {\em J. Math. Phys.} {\bf 21} 2386
\item Nagel B 1995 {\em Modern Group Theoretical Methods in 
      Physics}, ed J Bertrand {\em et al} (Dordrecht: Kluwer) p 211
\item Niederer U 1973 {\em Helv. Phys. Acta} {\bf 46} 191
\item Perelomov A M 1972 {\em Commun. Math. Phys.} {\bf 26} 222
\item ------ 1977 {\em Usp. Fiz. Nauk} {\bf 123} 23
\item ------ 1986 {\em Generalized Coherent States and Their 
      Applications} (Berlin: Springer)
\item Prakash G S and Agarwal G S 1995 {\em Phys. Rev.} A 
      {\bf 52} 2335
\item Segal I E 1962 {\em Illin. J. Math.} {\bf 6} 500
\item Sharma J K, Mehta C L and Sudarshan E C G 1978 {\em
      J. Math. Phys.} {\bf 19} 2089
\item Sharma J K, Mehta C L, Mukunda N and Sudarshan E C G 1981
      {\em J. Math. Phys.} {\bf 22} 78 
\item Solomon A I 1971 {\em J. Math. Phys.} {\bf 12} 390
\item Stoler D 1970 {\em Phys. Rev.} D {\bf 1} 3217
\item ------ 1971 {\em Phys. Rev.} D {\bf 4} 2308
\item Sudarshan E C G 1993 {\em Int. J. Theor. Phys.} {\bf 32} 1069
\item Trifonov D A 1994 {\em J. Math. Phys.} {\bf 35} 2297
\item Vilenkin N Ya 1968 {\em Special Functions and the Theory of
      Group Representations} (Providence, RI: Am. Math. Soc.) ch VII 
\item Vourdas A 1992 {\em Phys. Rev.} A  {\bf 45} 1943 
\item ------ 1993a {\em Phys. Scr.} T {\bf 48} 84
\item ------ 1993b {\em J. Math. Phys.} {\bf 34} 1223
\item Weyl H 1950 {\em The Theory of Groups and Quantum Mechanics}
      (New York: Dover)
\item W\"{u}nsche A 1991 {\em Quantum Opt.} {\bf 3} 359
\item ------ 1992 {\em Ann. der Phys.} {\bf 1} 181
\item ------ 1995 {\em Acta Phys. Slov.} {\bf 45} 413
\item Yuen H P 1976 {\em Phys. Rev.} A {\bf 13} 2226
\item Zhang W-M, Feng D H and Gilmore R 1990 {\em Rev. Mod. Phys.}
      {\bf 62} 867
\end{description}

\end{document}